\documentclass{ws-procs9x6}

\def\E{E}

\def\A{A}
\def\M{M}
\def\bi{\begin{itemize}}
\def\ei{\end{itemize}}
\newcommand{\beq}{\begin{equation}} 
\newcommand{\eeq}{\end{equation}}
\newcommand{\bea}{\begin{eqnarray*}} 
\newcommand{\eea}{\end{eqnarray*}}

\begin{document}


\title{Gravitation and cosmology in brane-worlds}

\author{David Langlois}
\address{GRECO, Institut d'Astrophysique de Paris, \\ 
98bis Boulevard Arago, 75014 Paris, France\\
email:langlois@iap.fr}  

\def\C{C}
\def\T{T^{\rm (bulk)}}
\def\M{M}
\newcommand{\gsim}{\ \raise.3ex\hbox{$>$\kern-.75em\lower1ex\hbox{$\sim$}} \ } 
\newcommand{\lsim}{\ \raise.3ex\hbox{$<$\kern-.75em\lower1ex\hbox{$\sim$}} \ } 


\maketitle

\abstracts{In brane-worlds, our universe is assumed to be 
a submanifold, or brane, embedded in a higher-dimensional 
bulk spacetime. Focusing on scenarios with a curved five-dimensional 
bulk spacetime, I discuss their gravitational and cosmological properties. 
}


\section{Introduction}
In this contribution, I review the basic ingredients of the so-called 
brane-world models. These  models 
are based on the  assumption that our universe is  
a {\it brane}: a sub-space embedded in a higher dimensional {\it bulk}
 spacetime. 
In contrast with the traditional Kaluza-Klein 
treatment  of extra dimensions, ordinary matter fields are {\it confined} 
on the brane. As a consequence, the size of the extra-dimensions 
in brane-world models is  not limited by the usual bound, typically 
$R\lesssim (1{\rm TeV})^{-1}$, based on the non-detection of Kaluza-Klein
modes in collider experiments up to energies close to $1$ TeV. In 
brane-worlds, only gravity propagates in the higher-dimensional 
spacetime and, therefore, 
the size of the extra-dimensions can be much larger, up to 
a fraction of a millimeter, corresponding to the current limit of 
gravity experiments that look for deviations of 
Newton's law\cite{Hoyle_etal04}.

Although there were some precursor works on brane-worlds, the huge 
interest they have encountered in the last few years is due  
to recent developments in string/M-theory. This has in turn opened up 
new avenues to tackle  some fundamental problems. 

For example, the fact that the 
four-dimensional Planck mass $M_p$ is in this context only a ``projection'' 
of the higher-dimensional (fundamental) Planck mass $M_*$, which can thus be 
lower than $M_p$, offers a new perspective on the hierarchy 
problem  and suggests the possibility
that quantum gravity might be  closer than previously thought.
This has been emphasized by
 Arkani-Hamed, Dimopoulos and Dvali\cite{add}, who
 suggested a very simple model
based  on a brane embedded in a 
flat geometry with $4+n$ dimensions, where $n$ dimensions are 
compactified on a torus of size $R$. Using Gauss' theorem, one can easily
show that, on scales $r\gg R$,  standard 4D gravity is recovered 
with $M_P^2=M_*^{2+n}R^n$. On scales $r\ll R$,
gravity becomes $(4+n)$-dimensional.

Another important progress was made by Randall and Sundrum\cite{rs99a,rs99b}, 
who 
considered  {\it curved}, or {\it warped}, 
bulk geometries. They showed, in particular, 
that  {\it compact} extra-dimensions are not necessary  to obtain a 
four-dimensional behaviour.
The bulk curvature can indeed lead to an 
{\it effective compactification}. 

 In this contribution, I will  focus my attention on 
 brane-worlds characterized by 
 a single extra  dimension   where  the bulk space-time is 
{\it curved} instead of flat and where 
the  self-gravity of the brane is  taken into account. This includes the 
configurations discussed by Randall and Sundrum.
  
The outline is the following. In the next section, I will  present  
in some detail the Randall-Sundrum 
models and discuss the corresponding effective gravity on the brane. 
Then, in the subsequent sections, I will turn my attention to cosmological 
brane models, reviewing a few important topics in brane cosmology.
In section 3, the bulk and brane geometries will be presented. 
Section 4 will be devoted to the so-called dark radiation, or Weyl radiation.
Inflation in the brane will be discussed in section 5. Section 6 will
discuss some aspects  
of brane cosmology when  the assumption of isotropy or homogeneity
is relaxed. 
Finally, section 7 will provide some conclusions and perspectives.

Only some aspects of brane-worlds are discussed in this contribution. 
Much more can be learnt on this subject  from several detailed 
 reviews~\cite{reviews}.

\section{Gravity in brane-worlds}
With extra-dimensions, gravity is intrinsically 
different from 4D gravity and the models must be able to mimic  standard 
gravity in the regimes tested by experiments. 
 The standard approach is to compactify the extra-dimensions 
on a size smaller than the scales probed by gravity experiments. 
However, as shown by Randall and Sundrum a few years ago, 
the extra-dimension can remain infinite,
the  ``effective compactification'' being the consequence of  
 the bulk spacetime curvature.

\subsection{The Randall-Sundrum model}
The (second) Randall-Sundrum\cite{rs99b}
model is based on the following ingredients
\bi
\item  a five-dimensional
bulk spacetime,  empty, but endowed with a negative cosmological constant
\beq
\Lambda=-{6\over\ell^2},
\eeq
\item a self-gravitating brane, which represents our world, 
endowed with a tension $\sigma$, 
and assumed to be $	Z_2$-symmetric. 
\ei
The five-dimensional Einstein equations are given by 
\beq
G_{AB}+\Lambda g_{AB}=\kappa^2 T_{AB}, 
\eeq
where $\kappa^2$ is the gravitational  coupling, and the corresponding
five-dimensional Planck mass, $M_5$, is defined by
\beq
  \kappa^2=M_5^{-3}.
\eeq
The bulk being empty, only the brane contributes to the energy-momentum
tensor $T_{AB}$. 
There are two equivalent ways of  solving Einstein's equation. Either
one solves it directly by taking into account the presence of the
brane, assumed to be infinitely thin along the extra-dimension, 
in the form of a {\it distributional} energy-momentum tensor. Or, one 
solves first 
 the {\it vacuum} Einstein equations, i.e. setting the right hand side 
to zero, and {\it afterwards}, one  takes into account the brane by imposing  
appropriate {\it junction conditions} at the spacetime boundary 
where the brane is located. These boundary  conditions are the 
generalization, to five dimensions, of the so-called Israel (-Darmois)
junction conditions and read 
\beq
[K_{AB}]=-\kappa^2\left(T_{AB}-{T\over 3}h_{AB}\right).
\eeq
They relate the jump, between the two sides of the brane, of 
 the  extrinsic curvature tensor, defined by 
$K_{AB}\equiv h_A^C{D_C n_B}$ (where $n^A$ is the unit vector normal to the
brane and $h_{AB}=g_{AB}-n_An_B$ is the induced metric on the brane), to 
the brane energy-momentum 
tensor. For a $Z_2$ symmetric brane, the jump of the extrinsic curvature
is simply twice the value of the extrinsic curvature on one side of the 
brane.

Provided the tension satisfies the {\it constraint} 
\beq
\label{rs_constraint}
\Lambda+{\kappa^4\over 6}\sigma ^2=0,
\eeq
which implies in particular that $\sigma= 6\M^3/\ell$, it can be 
shown that the five-dimensional Einstein equations admit  the 
following {\it static} solution
\beq
\label{rs}
ds^2=a^2(y)\eta_{\mu\nu} dx^\mu dx^\nu+dy^2,
\eeq
\begin{figure}[t]
  \begin{center}
    \includegraphics[height=10pc,angle=0]{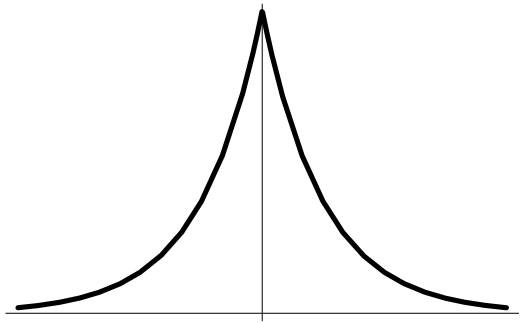}
  \end{center}
\caption{Warping factor $a(y)$.}
\end{figure}
where $\eta_{\mu\nu}$ is the usual Minkowski metric and 
$a(y)$ is a {\it warping} scale factor, whose explicit dependence 
on $y$ is given by
\beq
\label{a}
a(y)=e^{-|y|/\ell},
\eeq
as shown on Fig.~1. 
Here, the brane is located at $y=0$ and the $Z_2$ symmetry means
that  $y$ with $-y$ are identified.
This bulk solution (\ref{rs}-\ref{a}) 
can also be interpreted as two identical portions 
of AdS (Anti-de Sitter) spacetime glued together at the brane location.
 
\subsection{Gravity in the Randall-Sundrum model}
Let us now investigate the effective gravity for this 
model, as measured by  an observer located 
on the brane. 
A first, and rather simple, step is to compute the effective four-dimensional
Planck mass. This can be done by substituting in the five-dimensional 
Einstein-Hilbert action,
\beq
 S_{\rm grav}={\M^3\over 2}\int d^4x \, dy \, \sqrt{-g}\, R, 
\eeq
the metric (\ref{rs}) and by integrating over the extra-dimension. The 
factor in front of the resulting four-dimensional 
Einstein-Hilbert action (for $\eta_{\mu\nu}$)  gives 
\beq
\label{planck}
M_{Pl}^2=\M^3 \int_{-\infty}^{+\infty} dy\  a^2(y)= \M^3\ell.
\eeq
It is important to emphasize that the extra-dimension extends here to 
infinity. In the absence of the warping factor $a(y)$ this would lead 
to an infinite four-dimensional Planck mass. The warping of the 
extra-dimension, governed by the AdS lengthscale $\ell$, thus leads to 
an {\it effective compactification}, even if the extra-dimension is 
infinite.

To explore further the gravitational behaviour and derive 
for example the effective potential of a point mass located on the brane, 
one must study the  perturbations about the background metric 
(\ref{rs}).
Perturbing the metric, $g_{AB}=\bar g_{AB}+{h_{AB}}$, 
and working in the gauge $h_{yy}=0$, $h_{y\mu}=0$, $h_\mu^\mu=0$, 
$\partial_\mu h^\mu_\nu=0$, one finds that the linearized Einstein 
equations reduce to 
\beq
\left(a^{-2}\partial^2_{(4)}+\partial_y^2-{4\over \ell^2}+{4\over \ell}
\delta(y)\right)h_{\mu\nu}=0.
\eeq
This equation is separable and the solutions can be written as 
the superposition of eigenmodes $h(x^\mu,y)=u_m(y)
e^{ip_\mu x^\mu}$, with $p_\mu p^\mu=-m^2$. This implies that 
the dependence on the fifth dimension of the massive modes is governed 
by  the {\it Schr\"odinger-like equation}:
\begin{figure}[t]
  \begin{center}
    \includegraphics[height=12pc,angle=0]{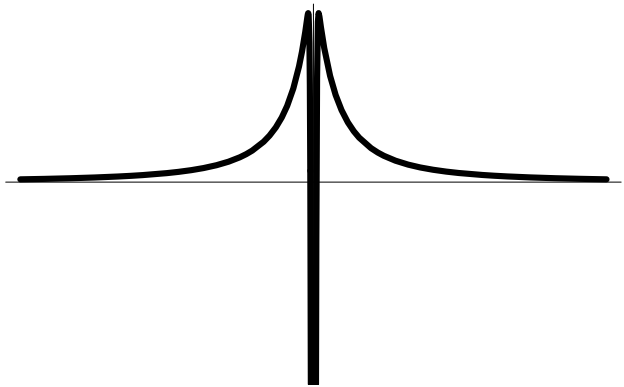}
  \end{center}
\caption{Effective potential of the Schr\"odinger-like equation that governs the dependence on the fifth dimension.}
\end{figure}
\beq
{d^2{\psi}_m\over dz^2}-V(z){\psi}_m=-m^2{\psi}_m,
\qquad
V(z)={15\over 4(|z|+\ell)^2}-{3\over \ell}\delta(z),
\eeq
using the function $\psi_m=a^{-1/2}u_m$ and the variable $z=\int dy/a(y)$.
The potential $V(z)$, plotted in Fig.~2, 
 is ``volcano''-shaped and 
goes to zero at infinity.

One can divide the 
solutions of this Schr\"odinger-like equation into: 
\bi
\item a zero mode ($m=0$),
$u_0(y)=a^2(y)/\sqrt{\ell}$,
which is concentrated near the brane and reproduces  the usual 
behaviour of 4D gravity;
\item a continuum of massive modes ($m>0$), which are weakly coupled to the 
brane and bring modifications with respect to standard 4D gravity.
\ei
More specifically, the perturbed metric 
outside a spherical source of mass $M$, and for $r\gg \ell$, is given by
\cite{gt} 
\beq
{\bar h}_{00}\simeq {2GM\over r}\left(1+{2\ell^2\over 3r^2}\right),\quad 
{\bar h}_{ij}\simeq {2GM\over r}\left(1+{\ell^2\over 3r^2}\right),
\eeq
where the bar here means that the perturbations have been  rewritten in 
a Gaussian
Normal gauge (i.e. $h_{yy}=h_{y\mu}=0$ and the brane is located at $y=0$) and 
thus correspond directly to the quantities measured on the brane. 
Standard gravity is thus recovered on scales $r\gg \ell$ !

On scales of the order of $\ell$, and below, one expects  deviations 
from the usual Newton's law.
Since  gravity experiments\cite{Hoyle_etal04}
 have  confirmed the standard Newton's 
law 
down to scales of the order $0.1$ mm, this implies  
\beq
\ell\lesssim 0.1 {\rm mm},
\eeq
and thus $M_{(5)}\gsim 10^8$ GeV.

Although 
the above results  apply to  linearized gravity, 
other works, based on second order calculations or numerical gravity,
have confirmed   the recovery 
of standard gravity on scales larger than $\ell$.
However, the behaviour of black holes in  the Randall-Sundrum model 
 might significantly 
deviate from the standard picture. Indeed, 
inspired by the AdS/CFT correspondence, it has been conjectured that 
Randall-Sundrum black holes  should {\it evaporate classically}, 
or, in other words,  be classically unstable.
The underlying argument is that the five-dimensional classical solutions 
should correspond to {\it quantum-corrected} 
four-dimensional black hole solutions, 
of a  conformal field theory (CFT) coupled to gravity\cite{tanaka,efk}. 
Since there are many  CFT degrees of freedom 
 into which the black hole can radiate, its 
 life time is shorter than for a standard black hole: 
\beq
\tau \simeq 10^2\left({M/ M_\odot}\right)^3\left(\ell/1{\rm mm}\right)^{-2}
{\rm years}.
\eeq

\subsection{Two-brane models}
Although I have been considering so far  a {\it single}
 brane (this  will also 
be the case for most of the discussion on cosmology), it is also worth 
mentioning models with {\it two} $Z_2$ symmetric branes. Because of the 
$Z_2$ symmetry, the two branes can  be seen as ``end-of-the-world'' branes
and therefore the extra-dimension is explicitly compactified.

In their first model\cite{rs99a}, Randall and Sundrum have introduced two such 
branes,  with opposite tensions $\sigma_\pm=\pm 6\M^3/\ell$,
separated by a distance $d$. Because both branes satisfy the 
constraint (\ref{rs_constraint}), one can still get a static configuration.
However, as far as gravity is concerned, 
there is a crucial difference between the  two branes. The 
effective gravity for a brane observer, 
ignoring the corrections 
due  the massive modes, is described  by a {\it scalar-tensor} theory 
with the Brans-Dicke parameter\cite{gt}
\beq
\omega_{\rm BD}^{(\pm)}={3\over 2}\left(e^{\pm 2d/\ell}-1\right),
\eeq
where the sign depends on which brane 
 the observer is located  (positive or negative 
tension respectively).
If we live on the positive tension brane, this is compatible 
with the present  constraint, $\omega_{\rm BD}>4\times 10^5$, 
provided the distance 
$d$ between the two branes is large enough.
The corresponding effective action is given by\cite{ks}
\beq
S_{\rm eff}=\int d^4x \sqrt{-g}\left[{1\over 2}\Psi R - {\omega(\Psi)\over 
2\Psi}\partial^\mu\Psi\partial_\mu\Psi- \sigma_+-\sigma_-
\left(1-\Psi\right)^2\right], 
\eeq
with 
\beq
\omega(\Psi)={3\over 2}{\Psi\over 1-\Psi}.
\eeq
However, if we live on the negative tension
brane, as was originally assumed by Randall and Sundrum in order to
solve the hierarchy problem, then the effective gravity is incompatible
with reality. To rescue this scenario, one must invoke a mechanism 
that stabilizes the distance between the two branes, e.g. by introducing 
a bulk scalar field coupled to the branes\cite{gw}.

\section{Homogeneous brane cosmology}
I now discuss the cosmology of a brane embedded in a five-dimensional 
bulk spacetime.

\subsection{The model}
As in standard cosmology, {\it homogeneity}
 and {\it isotropy} are assumed 
along  the three ordinary spatial dimensions. One thus requires 
the bulk spacetime to satisfy  the {\it cosmological symmetry}, 
which means that one can foliate the bulk 
 with maximally symmetric three-dimensional surfaces. This is 
in complete   analogy with {\it the spherical symmetry}, associated with 
(positively curved) maximally  symmetric two-dimensional surfaces
in a 4D spacetime. 

In addition to the three ordinary spatial dimensions, spanning the 
homogeneous and isotropic surfaces, one  introduces a time 
coordinate $t$ and a spatial coordinate $y$ for the extra dimension. The 
cosmological symmetry implies that the metric components depend only 
 on $t$ and $y$.
It is convenient to work in  a Gaussian Normal (GN)
 coordinate system, in which the brane is always located at $y=0$ and 
the five-dimensional metric takes   the form 
\begin{equation}
\label{GN}
ds^2=- n^2(t,y)dt^2+a^2(t,y)d\Sigma_k^2+dy^2,
\end{equation}
where $d\Sigma_k^2$ is the metric for the maximally symmetric 
three-surface ($k=0,\pm 1$). 
Note that, in closer analogy with the spherical symmetry mentioned above, 
another possibility would be to choose a coordinate system where 
the metric reads 
\beq
\label{spherical}
ds^2=- n^2(t,r)dt^2+b^2(t,r) dr^2+ r^2 d\Sigma_k^2.
\eeq

To obtain the equations governing the 
cosmological evolution, one substitutes the ansatz (\ref{GN}) into
 the five-dimensional Einstein equations
\begin{equation}
G_{AB}+\Lambda g_{AB}=\kappa^2 T_{AB}
\end{equation}
where  the energy-momentum tensor, assuming a  bulk otherwise empty, 
is due to the brane matter and  thus given by
\begin{equation}
T_A^B=Diag(-\rho_b(t), P_b(t), P_b(t), P_b(t), 0)\delta(y),
\end{equation}
where $\rho_b$ and $P_b$ are respectively the total energy density and 
pressure in the brane. 
The five-dimensional Einstein's 
equations can be solved explicitly~\cite{bdel99} and 
one  gets a solution for the metric components 
$n(t,y)$ and $a(t,y)$, in terms of $\rho_b(t)$ and $P_b(t)$, defined 
up to an integration constant. 

\subsection{The cosmological evolution in the brane}
On the brane, the metric is given by 
\beq
ds_b^2=-n_b(t)^2 dt^2 +a_b(t)^2 d\Sigma_k^2,
\qquad n_b(t)\equiv n(t,0), \quad a_b(t)\equiv a(t,0).
\eeq
It can be shown that the scale factor $a_b(t)$ satisfies  
the {\it modified 
Friedmann equation}~\cite{bdl99,bdel99}:
\begin{equation}
\label{fried}
H_b^2\equiv {\dot a_b^2\over a_b^2}={\Lambda\over 6}+{\kappa^4\over 36}
\rho_b^2+{\C\over a_b^4}-{k\over a_b^2},
\end{equation} 
where $\C$ is an integration constant.
It can also be shown that, for an empty bulk, the  
usual  conservation equation  holds, which implies
\begin{equation}
\dot\rho_b+3H_b(\rho_b+P_b)=0.
\end{equation}

 For $\Lambda=0$ and $\C=0$,  the bulk is 5-D Minkowski and  the cosmology 
is highly unconventional since the Hubble parameter is proportional to 
the brane energy density~\cite{bdl99}. This has the 
unfortunate   consequence  to 
ruin the standard nucleosynthesis scenario, which is based 
 on the evolution of
the expansion rate with respect to the relevant microphysical 
interaction rates. 

To obtain a viable brane cosmology scenario, the simplest way is to 
generalize the  Randall-Sundrum model
to cosmology\cite{cosmors}. In the 
static version of the previous section, the energy density of the 
``Minkowski'' brane 
is  $\rho_b=\sigma_{RS}\equiv 6\M^3/\ell $. This can be generalized to 
a ``FLRW'' brane by adding to the intrinsic tension $\sigma_{RS}$ 
the usual cosmological energy density
$\rho(t)$ so that the total energy density is given by 
\beq
\label{rho_b}
\rho_b(t)=\sigma_{RS}+\rho(t).
\eeq
Moreover, the bulk is assumed to be endowed 
 with a negative cosmological constant $\Lambda<0$, satisfying the 
constraint (\ref{rs_constraint}).

Substituting the decomposition (\ref{rho_b}) into
 the Friedmann equation  (\ref{fried}), one finds 
\begin{equation}
\label{bdl}
H_b^2={8\pi G\over 3} \rho +{\kappa^4\over 36}\rho^2
+{\C\over a_b^4}-{k\over a_b^2}.
\end{equation}
In the expansion in $\rho$, the constant term vanishes because of the 
constraint (\ref{rs_constraint}), whereas the coefficient of the linear 
term is the standard one because 
$
8\pi G\equiv \kappa^4\sigma/6$, as implied by 
 (\ref{rs_constraint}) and (\ref{planck}).
However, the Friedmann equation (\ref{bdl}) 
 is characterized by two new features:
\begin{itemize}
\item a $\rho^2$ term, which dominates at high
energy;
\item a radiation-like term,  ${\C/ a_b^4}$, usually called 
{\it dark radiation}.
\end{itemize}
The cosmological evolution undergoes  a transition from a high 
energy regime, $\rho\gg \sigma$, characterized by 
an unconventional behaviour of the scale factor, into a 
low energy regime which reproduces our standard cosmology. For 
$\C=0$, $k=0$ and an equation of state $w=P/\rho=const$, one can solve 
analytically the evolution equations and one finds
\begin{equation}
a(t)\propto t^{1/q}\left(1+{q\, t\over 2\ell}\right)^{1/q},
\qquad q=3(1+w).
\end{equation}
One clearly sees the transition, at the epoch $t\sim \ell$, between 
the early, unconventional, evolution $a\sim t^{1/q}$ and the standard
evolution $a\sim t^{2/q}$.

In order to 
be compatible with the nucleosynthesis scenario,  
the high energy regime, where the cosmological evolution is unconventional, 
must take place before 
nucleosynthesis. This requires $\sigma^{1/4} \gsim 1$ MeV, and since 
$\sigma= 6/(\kappa^2\ell)=6 M_5^6/M_P^2$, this gives the constraint
$M_5\gsim 10^4$ GeV. One notes that this 
is much less stringent than the constraint 
from   small-scale gravity experiments, 
which presently require $\ell \lsim 0.1$ mm and $M_5 \gsim 10^8$ GeV.
As will be detailed in the next section, another observational constraint
applies to the dark radiation constant $\C$.

\subsection{Another point of view}
If,  instead of the GN ansatz (\ref{GN}) for the metric, 
one starts from the metric (\ref{spherical}), in analogy with 
the spherical symmetry, one recognizes the generalization of the Birkhoff 
theorem, which states that a vacuum spherical symmetric solution of 
Einstein's equation is necessarily static and gives the Schwarschild 
metric: the 5D vacuum cosmologically symmetric solution of 5D Einstein's 
equations with a (negative) cosmological constant is necessarily 
static and corresponds to the 
 AdS-Schwarzschild metric in five dimensions:
\begin{equation}
\label{ads}
ds^2=-f(R)dT^2+{dR^2\over f(R)}+R^2 d\Sigma_k^2,
\quad
f(R)=k+{R^2\over \ell^2}-{\C\over R^2}, \quad k=0,\pm 1.
\end{equation}
In this coordinate system, the brane is {\it moving} and the so-called 
junction conditions 
$[K_{\mu\nu}]=-\kappa^2\left(S_{\mu\nu}-(S/3)g_{\mu\nu}\right)$
give the modified Friedmann equation obtained above~\cite{kraus,ida}.

\section{Dark radiation}

So far, the bulk has been assumed to be {\it strictly empty}, apart from 
the presence of the brane.
However, the fluctuations of brane matter generate bulk 
gravitational waves. Equivalently, the scattering of brane 
particles  produce bulk gravitons 
($
\psi+{\bar \psi}\rightarrow G
$). 
Therefore, a realistic model must take into account the presence
of these bulk gravitons, 
which are emitted by the brane and then propagate in 
the bulk.

\subsection{Emission and propagation of bulk gravitons}
The rate of emission of these gravitons by the brane can be computed 
explicitly 
when the brane matter is in thermal equilibrium (with a temperature 
$T$). The corresponding energy loss rate is given by\cite{lsr02,hm01} 
\begin{equation}
\label{emission}
\dot\rho+4H\rho=-{315\, \over 512\, \pi^3}
\, \hat g \, \kappa^2\,  T^8,
\end{equation}
with the effective number of degrees of freedom
\beq
\hat g=(2/3)g_s+4g_v +g_f,
\eeq
which is weighted sum of the scalar, vector and fermionic degrees of 
freedom.

After their emission, the gravitons propagate freely in the bulk 
where they follow geodesic trajectories. As illustrated in Fig.~3, some 
of these 
gravitons (in fact many)  tend to come back onto 
the brane and  bounce off it.
\begin{figure}[t]
  \begin{center}
    \includegraphics[height=13pc]{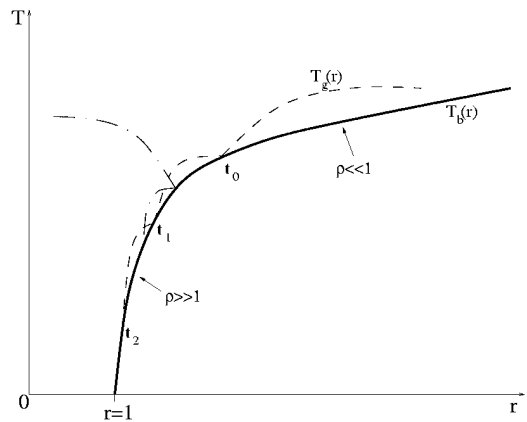}
  \end{center}
  \caption{Examples of graviton trajectories in the bulk (dashed line). 
The brane 
trajectory, with relativistic matter, is also shown (continuous line).}
\end{figure}
All these gravitons contribute to an effective bulk energy-momentum
tensor, which can be written as 
\begin{equation} 
\T_{AB}=\int d^5p \ \delta\left(p_Mp^M\right)\sqrt{-g}\,f\, p_Ap_B, 
\label{T_integ},
\end{equation}  
where $f$ is the phase space distribution function.

 From the 5D Einstein equations,
one can derive effective 4D Einstein equations~\cite{sms99}, 
which in the homogeneous 
case yield 
\begin{itemize}
\item the  Friedmann equation
\begin{equation}
H^2 =   
{8\pi G \over 3}\left[  
\left(1+{\rho\over 2\sigma}\right)\rho+ \rho_{\rm{D}}  \right],  
\label{Hubble} 
\end{equation} 
\item the non-conservation equation for brane matter, 
which must be identified with (\ref{emission}),
\begin{equation} 
\dot{\rho}+3\,H\,\left(\rho+p\right)=2\,{\T}_{RS}\,n^R\,u^S\, ,
\end{equation} 
where $n^A$ is the unit vector normal to the  brane and 
$u^A$ its velocity in the bulk;
\item the non-conservation equation for the ``dark'' component
$\rho_{\rm{D}}$ (which includes all effective contributions from the bulk):
\begin{equation}
\label{dark}
 \dot \rho_{\rm{D}}+4H\rho_{\rm{D}} 
  =-2\left(1+{\rho\over\sigma}\right)\T_{AB} u^A n^B   
  -2 H\ell \, \T_{AB} n^A n^B\,. 
\end{equation} 
\end{itemize}
On the right hand side of this last equation, we find two terms 
involving the bulk energy-momentum tensor: 
the first term, due to the energy flux from the brane 
into the bulk, contributes positively and thus increases the amount of 
dark radiation whereas the second term, due to the pressure along 
the fifth dimension, decreases the amount of dark radiation.
These terms can be estimated numerically~\cite{ls03}. A striking 
property is  that the 
 gravitons  coming  back onto 
the brane and  bouncing off it  give a significant contribution 
to the 
transverse pressure effect, which almost, although not quite, compensates
the flux effect. 
The evolution of the dark radiation, or rather its ratio with respect 
to the brane radiation density
$\epsilon_D\equiv \rho_D/\rho$,  is plotted on Fig.~4.
\begin{figure}[t]
  \begin{center}
    \includegraphics[height=20pc,angle=-90]{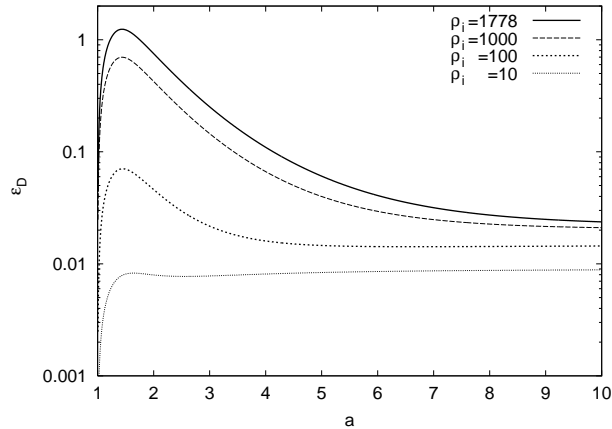}
  \end{center}
  \caption{Evolution of the  ratio  $\epsilon_D=\rho_D/\rho$ for different 
values of the initial energy density on the brane 
$\rho_i$ (in units of $\sigma^4$).}
\end{figure}
One observes that at late times, i.e. far in the low energy regime, the 
ratio reaches a plateau, because the right hand side of (\ref{dark}) becomes
negligible. The dark component then scales exactly like radiation.

\subsection{Observational constraints}
The computed amount of dark radiation can be confronted to observations. 
Indeed, since 
dark radiation behaves as radiation, it must satisfy the  
 nucleosynthesis  constraint on the number of {\it additional 
relativistic degrees of freedom}, usually expressed in terms 
of the extra number of light neutrinos $\Delta N_\nu$.
 The relation between $\Delta N_\nu$ and $\epsilon_D$ is given by
\begin{equation}
\epsilon_D={7\over 43}\left({g_*\over g_*^{\rm nucl}}\right)^{1/3}
\Delta N_\nu,
\end{equation}
where $g_*^{\rm nucl}=10.75$ is the number of degrees of freedom at
nucleosynthesis (in fact before the electron-positron annihilation).
Assuming $g_*=106.75$ (standard model), this gives
$\epsilon_D\simeq 0.35 \Delta N_\nu$. 
The typical constraint from nucleosynthesis 
$\Delta N_\nu \lesssim 0.2$
thus implies
\begin{equation} 
\epsilon_D\equiv {\rho_D\over\rho_r} \lesssim 0.03 \left({g_*\over 
g_{*}^{\rm nucl}}
\right)^{1/3},
\end{equation}
which gives $\epsilon_D \lesssim 0.09$ with the degrees of freedom 
 of the standard model.

\section{Brane inflation}
In brane cosmology,  the famous horizon problem  is much less 
severe than in standard cosmology, 
because the gravitational horizon, associated with the signal propagation
in the bulk, can be much 
bigger than the standard photon horizon, associated with the 
signal propagation on the brane\cite{cl01}.
However, it  is still alive
because the 
energy density on the brane is limited by the  Planck limit $\rho\sim M^4$.
Thus, one must still invoke inflation, altough alternative ideas 
based on the collision of branes\cite{kost}
 have been actively  explored (however, the 
generation of 
a quasi-scale-invariant fluctuation spectrum, as required 
by observations, remains problematic).

The simplest way 
to get inflation in the brane is to detune the brane tension from 
its Randall-Sundrum value (\ref{rs}) in order to obtain  a 
net effective four-dimensional cosmological constant that is positive. This
 leads to  exponential expansion on the brane.
In the GN coordinate system, the metric 
is separable and  
can be written as 
\beq
ds^2= \A(y)^2 \left(-dt^2+e^{2H t} d{\vec x}^2\right)+dy^2,
\label{dS}
\eeq
with 
\beq
\A(y)=   \cosh\mu y-\left(1+{\rho\over\sigma}\right) \sinh\mu|y|.
\label{A}
\eeq

As in the Randall-Sundrum case, the linearized Einstein equations 
for the tensor modes lead to a {\it separable} wave equation.
The shape along the fifth dimension 
of the corresponding massive modes is governed by 
the Schr\"odinger-like equation 
\beq
\label{SE}
 {d^2\Psi_m\over dz^2} - V(z)\Psi_m =-m^2 \Psi_m \,,
\end{equation}
after introducing   the new variable  
$z-z_b=\int_0^y d\tilde y/\A(\tilde y)$ 
(with $z_b=H^{-1}\sinh^{-1}(H\ell)$)
and the new function
 $\Psi_m= \A^{-1/2}u_m(y)$.
The potential is given by 
\beq
V(z)= {15H^2 \over 4\sinh^2(H z)} +
{{9\over4}}H^2
- {3\over\ell}\left(1+{\rho\over\sigma}\right) \delta(z-z_{\rm b}) \,
\eeq
and plotted in Fig.~5. 
\begin{figure}[t]
  \begin{center}
    \includegraphics[height=12pc,angle=0]{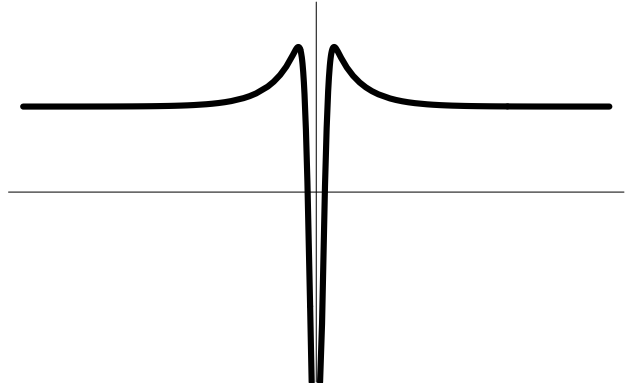}
  \end{center}
\caption{Potential for the graviton modes in a de Sitter brane}
\end{figure}
In constrast with the Randall-Sundrum potential, the potential 
goes asympotically to the non-zero value $9H^2/4$. This indicates 
 the presence 
of a gap
between the zero mode ($m=0$) and the continuum of Kaluza-Klein modes 
($m>3H/2$).

In practice, inflation is not strictly de Sitter but the de Sitter case 
discussed above is a good approximation when $\dot H\ll H^2$. To get 
``realistic'' inflation in the brane, two main approaches 
 have been considered: 
either to assume  a five-dimensional  scalar field which induces inflation
in the brane\cite{sasaki}, or to suppose  
a four-dimensional scalar confined on the
 brane\cite{mwbh99}.

In the latter case, the cosmological evolution during inflation 
is obtained by substituting the energy density 
$\rho_\phi=\dot\phi^2/2 + V(\phi)$
in the modified Friedmann equation (\ref{bdl}). 
For slow-roll inflation, this can be approximated by 
\beq
H^2\simeq {8\pi G\over 3}\left(1+{V\over 2\sigma}\right)V.
\eeq
Interestingly, because of the modified Friedmann equation, 
new features appear at  high energy ($V> \sigma$):
the slow-roll 
conditions are changed and,  because the Hubble parameter is bigger 
than the standard value, yielding a higher friction on the scalar field,
inflation can occur  with potentials usually too steep to 
sustain it~\cite{cll00}.

The scalar and tensor spectra generated during inflation driven by 
a brane scalar field have also been computed~\cite{mwbh99,lmw00}. 
They are modified 
with respect to the standard results:
\beq
P_S=P_S^{\rm(4D)}\left(1+{V\over 2\sigma}\right)^3,
\qquad
P_T=P_T^{\rm (4D)}F^2(H\ell),
\eeq
with 
\beq
F\!\left(x\right) =\left\{ \sqrt{1+x^2} - x^2 \ln \left[ {1\over
x}+\sqrt{1+{1\over x^2}} \right] \right\}^{\!\!-1/2}
\!.
\eeq
 $F\simeq 1$ at low energies, i.e. for $H\ell\ll 1$, whereas 
  $F\simeq {(3/2)H\ell}\sim V/\sigma$ at very high energies, 
i.e. for $H\ell\gg 1$.
At low energies, $F\simeq 1$ and 
one  recovers exactly the usual four-dimensional result 
but at higher energies the multiplicative factor $F$ provides an 
{\it enhancement} of the gravitational wave spectrum amplitude 
with respect to the four-dimensional result. However, comparing  
this with the amplitude for the scalar spectrum,
one finds that, at high energies ($\rho\gg\sigma$), the {\it 
tensor over scalar 
ratio is in fact suppressed} with respect to the four-dimensional ratio.
However, a
 difficult question is how the perturbations  will evolve during the 
subsequent cosmological phases, the radiation and matter eras.

\section{Beyond homogeneous brane cosmology}
The homogeneous and isotropic brane cosmology is in fact very 
simple because  of the generalized Birkhoff's theorem mentioned 
earlier. But, when the cosmological symmetry is relaxed, things 
become rather  difficult because the bulk geometry is no longer 
Schwarzschild-AdS. This section presents some aspects of this complexity, 
starting with anisotropic, but still homogeneous, models and then
discussing the evolution of the linear cosmological perturbations.

\subsection{Anisotropic brane cosmology}
Let us now discuss configurations where the  
 cosmology in the brane is homogeneous 
but anisotropic, e.g. of the Bianchi 
I type with a metric of the form
\beq
ds^2_b=-d\tau^2+\sum_{i=1}^3a_i^2(\tau)(dx^i)^2.
\eeq
Although many works in the literature 
have been devoted to this subject, most of them use  the effective
four-dimensional equations projected on the brane. It is a  more 
challenging task to solve the 5D Einstein equations for the bulk as well, 
starting e.g. from an ansatz of the form 
\beq
 ds_{\rm bulk}^2=-{n^2(t,y)}dt^2+\sum_{i=1}^3
{a_i^2(t,y)}({dx^i})^2+dy^2.
\eeq
Assuming that the metric is separable, it turns out that  
analytical  solutions can be obtained~\cite{flsz04}. The five-dimensional 
metric is given by  
\begin{eqnarray}
ds^2=\sinh^{1/2}(4y/\ell)\left[\right. &-& \tanh\left(2y/\ell\right)^{2 q_0} dt^2
\cr
&+&\left.\sum_i \tanh\left(2y/\ell\right)^{2 q_i}t^{2p_i}\left(dx^i\right)^2\right]
+dy^2,
\end{eqnarray}
where the seven coefficients $q_\mu$ and $p_i$ must satisfy the constraints
\beq
\sum_{\mu=0}^3 q_\mu=0, \quad \sum_\mu q_\mu^2={3\over 4}, \quad 
\sum_{i=1}^3 p_i=1 , 
\quad \sum_i p_i^2=1, \quad \sum_i q_i\left(p_i+1\right)=0.
\eeq

In general, a brane embedded in an anisotropic bulk spacetime must
contain matter with {\it anisotropic stress}, 
because of the junction conditions:
\bi
\item  isotropic part:
\beq
n^{-1}\dot y_b  \left. \dot{A} \right|_b + 
\sqrt{1+\dot y_b^2 } \, \left. {A'} \right|_b = 
\frac{\kappa^2}{6} \rho_b \label{Cond1},
\eeq
\item anisotropic part:
\beq
n^{-1}\dot y_b \left. \dot{B}_i \right|_b +
\sqrt{1+\dot y_b^2}\; \left. {B'}_i \right|_b =
\frac{\kappa^2}{2}\pi_i, \label{Cond2}
\eeq
where $\pi_i$ is the anisotropic pressure in the brane,
\ei
with the  notation $3A\equiv\ln(a_1a_2a_3)$ and 
$B_i\equiv\ln a_i -A$.
Note that the brane position $y_b$ is not assumed to be fixed 
here: in this sense
the coordinate system is not Gaussian Normal.
Interestingly, the above  solutions include 
a particular bulk geometry, for $q_0=\pm \sqrt{3}/4$, in which 
one can embed a moving 
brane with  perfect fluid as  matter.
The effective cosmological equation of state
$P_{\rm eff}/ \rho_{\rm eff}$ is negative but goes to zero 
at late times.

\subsection{Scalar cosmological perturbations}
A crucial step  for brane cosmology is to be  confronted with cosmological 
observations, in particular the  CMB fluctuations.  Although the {\it 
primordial}
power spectra for scalar and tensor perturbations have been computed, the 
subsequent evolution  of the cosmological perturbations is non trivial 
and has not been fully solved yet. The reason for this is that, 
in contrast with standard cosmology where the 
evolution of cosmological perturbations can be reduced to  ordinary 
differential equations for the Fourier modes, the evolution equations in brane
cosmology are partial differential equations with two variables: time 
and the fifth coordinate. 
Another delicate point is to specify the boundary conditions, both in time 
and space. 

An instructive, although limited, approach for the brane cosmological 
perturbations is the brane point of view, based on the 4D effective Einstein
equations on the brane, usually written in the form~\cite{sms99}
\beq
\label{einstein_4d}
G_{\mu\nu}+\Lambda_4 g_{\mu\nu}=8\pi G \tau_{\mu\nu}+\kappa^2 \Pi_{\mu\nu}
-E_{\mu\nu},
\eeq
where $\tau_{\mu\nu}$ is the brane energy-momentum tensor, 
$\Pi_{\mu\nu}$ is a tensor depending quadratically on  $\tau_{\mu\nu}$, 
giving in particular the $\rho^2$ term in the Friedmann equation, and 
$E_{\mu\nu}$, which corresponds to the dark radiation in the 
homogeneous case, is the projection on the brane of the bulk Weyl tensor.

It is then a straightforward exercise, starting from (\ref{einstein_4d}),
 to write  down explicitly the perturbed effective
Einstein equations on the brane, which will look exactly as the 
four-dimensional ones for the geometrical part but with extra terms 
due to $\Pi_{\mu\nu}$ and $T^{Weyl}_{\mu\nu}$ . One thus gets
 equations relating  the perturbations of the metric 
to  the matter  perturbations {\it and} the perturbations of 
the projected Weyl tensor, which formally can be assimilated to a virtual 
fluid, with  corresponding (perturbed) energy density $\rho_E+\delta\rho_E$,
pressure $P_E+\delta
P_E={1\over3}(\rho_E+\delta\rho_E)$ and anisotropic pressure. 
The contracted Bianchi identities ($\nabla_\mu G^\mu_\nu=0$)
and energy-momentum conservation for matter on the brane 
($\nabla_\mu \tau^\mu_\nu=0$) ensure, using Eq.~(\ref{einstein_4d}), that
\begin{equation}
\nabla_\mu E^\mu_\nu = \kappa^4\,\nabla_\mu\Pi^\mu_\nu \,.
\end{equation}
In the background, this tells us that $\rho_\E$ behaves like radiation, 
as we knew already,
and for the first-order perturbations, one finds
 that the effective energy of the projected Weyl
tensor is conserved independently of the quadratic energy-momentum
tensor. The only interaction is a momentum transfer.
 
It is also possible to construct~\cite{lmsw00} gauge-invariant variables 
corresponding to the curvature perturbation on hypersurfaces of
uniform density, both for the brane matter energy density 
and for the total effective energy density (including the quadratic terms 
and the Weyl component). These quantities are extremely useful because 
their evolution on scales larger than the Hubble radius can be solved easily.
However, their connection to the large-angle 
CMB anisotropies involves the knowledge of anisotropic stresses due to the 
bulk metric perturbations. 
This means that  {\it for a quantitative prediction 
of  the CMB anisotropies, even at large scales,  one needs to determine 
the evolution of the bulk perturbations}.

In summary, one can  obtain a set of equations for the brane linear 
perturbations,  
where one recognizes the ordinary cosmological  equations 
but modified by   two  types of corrections:
\begin{itemize}
\item modification of the  homogeneous background coefficients due to the 
additional $\rho^2$ terms in the Friedmann equation.   These  corrections are  
negligible in the low energy  regime $\rho\ll\sigma$;
\item presence of source terms in the  equations. 
These terms come from the bulk perturbations and cannot be determined solely 
from the evolution inside the brane. To determine them, one must solve 
the full problem in the bulk (which also means to specify some initial 
conditions in the bulk). 
\end{itemize}

One should also mention a very recent work~\cite{Ichiki:2004sx} 
on the post-inflation evolution
of gravitational waves, which indicates that, on very small scales, 
the spectral density of gravitational waves is {\it reduced} with 
respect to the standard 4D result.

 \section{Conclusions}
In this contribution, I have presented some aspects of brane-world 
models, covering both the (static) Randall-Sundrum model and its cosmological
extensions. Due to lack of time/space, I have not discussed many 
other  interesting topics in the field. Examples are the brane cosmology 
of  models involving 
Gauss-Bonnet corrections; the induced gravity  models, where 
one includes a 4D Einstein-Hilber action for the brane and which 
can lead to late-time cosmological effects mimicking dark energy.

There are still many open questions in brane cosmology. Even in the simplest 
set-up, discussed here, based on a cosmological extension of the 
Randall-Sundrum model, the evolution of cosmological perturbations has not
yet been solved, although some significant progress has been made.
 The situation is still more complicated  in more sophisticated models, 
involving a bulk scalar field and/or collision of branes. 
It must be emphasized that the predictions for the  cosmological 
perturbations, as observed in the CMB experiments, and their adequation 
with the present data, 
is a crucial test
for brane-world models for which the early universe is modified.
More direct tests of 
brane-world models involve 
 gravity experiments or  collider experiments. However, 
if the fundamental Planck mass is too high, such direct experiments
cannot see extra-dimensional effects and one must turn to cosmology to 
try to see indirect signatures from the early universe.

Another direction of  research is to make contact between the 
brane-worlds, which are still only  phenomenological models, 
and a fundamental theory like string theory.




\end{document}